\documentstyle[aps,prl,epsf,twocolumn]{revtex}
\begin{document}
\def\bea{\begin{eqnarray}}
\def\eea{\end{eqnarray}}
\def\a{\alpha}
\def\d{\delta}
\def\p{\partial}
\def\nn{\nonumber}
\def\r{\rho}
\def\rv{\bar{r}}
\def\la{\langle}
\def\ra{\rangle}
\def\e{\epsilon}
\def\o{\omega}
\def\n{\eta}
\def\g{\gamma}
\def\break#1{\pagebreak \vspace*{#1}}
\def\f{\frac}
\draft
\twocolumn[\hsize\textwidth\columnwidth\hsize\csname
@twocolumnfalse\endcsname
\title{DNA Elasticity : Topology of Self-Avoidance}
\author{Joseph Samuel, Supurna Sinha and Abhijit Ghosh}
\address{Harish Chandra Research Institute,\\
Chhatnag Road, Jhunsi, Allahabad 211 019, India\\
and\\
Raman Research Institute,\\
Bangalore 560080, India.\\}
\maketitle
\widetext
\begin{abstract}
We present a theoretical treatment of DNA stretching and twisting
experiments, in which we discuss global topological 
subtleties of self avoiding ribbons 
and provide an underlying justification for
the worm like rod chain (WLRC) model proposed by Bouchiat and Mezard.
Some theoretical points regarding the WLRC model are clarified:
the ``local writhe formula'' and the use of an adjustable cutoff 
parameter to ``regularise'' the model.
Our treatment brings out the precise relation between 
the worm like chain (WLC), the paraxial worm like chain (PWLC) and the 
WLRC models. 
We describe the phenomenon of ``topological untwisting''
and the resulting collapse of link sectors in the WLC model
and note that this leads to a free energy profile {\it{periodic}} 
in the applied link. This periodicity disappears when one
takes into account the topology of self avoidance or at large
stretch forces (paraxial limit). 
We note that the difficult nonlocal notion of self avoidance can
be replaced (in an approximation) by the simpler local notion
of ``south avoidance''. 
This gives an explanation for the efficacy
of the approach of Bouchiat and Mezard in explaining the ``hat curves''
using  the WLRC model, which is a south avoiding model.
We propose 
a new class of experiments 
to probe the continuous transition 
between the periodic and aperiodic behavior of the free energy. 
\end{abstract}
]
\narrowtext
\section{Introduction}

Motivated by experiments \cite{bust,strick} in which single DNA molecules
are stretched and twisted to measure their elastic properties, Bouchiat
and Mezard (BM) \cite{bouchiat} have proposed the Worm Like Rod Chain
(WLRC) model, which gives a fair fit to the experimental data. However,
several theoretical aspects of the WLRC model remain unclear, as evidenced
by recent letters in PRL \cite{maggs,mezreply}. 
Points of dispute are BM's use of
a local Fuller writhe formula (as opposed to the non-local
C\u{a}lug\u{a}reanu-White formula\cite{cal}) and the need for a 
phenomenologically
introduced cutoff parameter, which has to be adjusted to fit the data.

The theoretical issues raised by the experiments of \cite{strick} are
surprisingly subtle. A reading of Bouchiat and Mezard's discussion of the 
``local writhe formula''\cite{bouchiat} may give the misleading 
impression that the C\u{a}lug\u{a}reanu-White formula and the 
Fuller formula \cite{fuller,fain} are simply {\it different} ways of 
expressing 
the {\it same} quantity, the writhe. In fact, there can be 
no local writhe formula. The writhe which appears in 
C\u{a}lug\u{a}reanu's theorem $Lk = Tw + Wr$ is given by the non-local  
C\u{a}lug\u{a}reanu-White formula and jumps by two units when the 
curve is passed through itself. No local integral can replicate this 
behaviour. Bouchiat and Mezard use the ``local writhe formula'' under
the assumption that the Euler angles are regular functions of the 
arc length parameter $s$. This assumption breaks down at the coordinate
singularities of the Euler angles. So, Bouchiat and Mezard are 
effectively computing the distribution of the ``Fuller writhe'', which 
is {\it not} the same as the C\u{a}lug\u{a}reanu-White writhe. 
Nevertheless, Bouchiat and Mezard obtain good agreement with 
experiment. The question remains: Why does it work? 
The purpose of this paper is to clarify these issues:
we show that these questions are related to the {\it topology} of self
avoidance.

This paper is structured as follows: we first describe the
experimental setup and summarize the experimental data. We then explain
the problem that we address, which concerns a global topological subtlety
in the configurations of ribbons.  In resolving the problem we point out
the connection between the Worm Like Chain (WLC)\cite{footwlc}, the
Paraxial Worm Like Chain (PWLC) and the Worm Like Rod Chain (WLRC) models.
We propose an experiment which brings out the relation 
between the models by continuously interpolating
between them. We conclude the paper by comparing with previous work.
\section{Experiments} Micromanipulation techniques are now so
sophisticated that experimenters can stretch and twist single DNA
molecules to probe their elastic properties under a torsional constraint.
In a typical experiment\cite{strick}, one end of a single molecule of
double stranded DNA is attached to a glass plate and the other to a
magnetic bead. The glass plate is kept fixed and the bead is
pulled by magnetic field gradients and
rotated by magnetic fields. By such
techniques the molecule is stretched and twisted by hundreds of turns and
the extension of the molecule is measured as a function of the number of
turns applied to the bead for a fixed force \cite{strick}.  Another
experiment measures the torque-twist relation at fixed force\cite{bryant}
using a slightly different experimental technique.  The length of the DNA
molecule is typically about 20 $\mu m$, its thickness $2$ $nm$ and the
bead is about $4.5 \mu m$ in diameter. In practice, this size of bead is
adequate to prevent the molecule from untwisting by looping around the
bead \cite{maggs}. A typical experimental plot is shown in Ref.
\cite{strick}. These curves are easy to understand qualitatively:  as one
twists the molecule, its extension progressively decreases as one can
guess by playing with a cord or ribbon. Further twisting leads to buckling
and the formation of twisted braids or ``plectonemes'' which are familiar
on telephone cords. 
Electron micrographs \cite{vino} of DNA show branched 
polymeric structures which 
indicate the formation of plectonemes.

To understand the experimental curves quantitatively more work is needed.
Unlike telephone cords, DNA is seriously affected by 
thermal fluctuations and there
are entropic effects to be accounted for. Several 
papers have already treated this problem 
\cite{bouchiat,maggs,nelson,writhe}.
The high force regime, which is amenable to perturbation theory 
about the taut polymer 
configuration is well understood \cite{nelson,writhe}. 
In the low force 
regime where self-avoidance effects are appreciable\cite{kamien},  
a consensus is still lacking and this is the primary focus of this 
paper.

\section{Theoretical Models}
To define a theoretical model we need to specify the allowed 
configurations ${\bf C}$ of the polymer given the experimental 
constraints, write 
down a microscopic Hamiltonian or energy $E({\cal C})$ for each 
configuration ${\cal C} \epsilon {\bf C} $  and
express the partition function as  a sum over configurations with 
Boltzmann weight.   
All experimentally accessible quantities can be got from the partition 
function. The problem therefore is to calculate the partition function
\begin{equation}
Z({\vec r}, Lk) = \sum_{{\cal C} \epsilon {\bf C}}{e^ {-E({\cal C}) 
/k_{B}T}}
\label{part}
\end{equation}
where ${\vec r}$ is the vector separation between the ends of the 
molecule, $Lk$ 
is the number of times the bead has been turned 
(which could be fractional)
and the summation 
appearing in the expression for the partition function represents
a sum over all allowed configurations of the polymer. 
$E({\cal C})$ is an energy functional which assigns an energy 
$E({\cal C})$ to each allowed configuration ${\cal C}$. 
Specification of the allowed configurations and this  
energy functional 
defines the model completely. The next step is to ``solve''
the model, {\it i.e} deductively work out its experimental
consequences. The objective is to confront theory with
experiment and learn from the discrepancy as well as the agreement. 
Unfortunately, ``solving'' the model is 
not always a practical proposition even in idealised situations. 
Solvable theoretical models are therefore a valuable aid to understanding.

It is often convenient to deal with conjugate variables
$(B,{\vec f})$ instead of $(Lk,{\vec r})$ and the Fourier-Laplace
transform of the partition function.
${\tilde Z}(B,{\vec f}) = \int{d{Lk}\,d{\vec r}{Z(Lk, {\vec r})}
e^{-iBLk+{\vec 
f}\cdot{\vec 
r}} 
}$.
 We shall restrict ourselves to the limit of long polymers 
(i.e. for a polymer of contour length $L$ and persistence length
$L_P$, $L/L_P 
\rightarrow \infty $).
In this limit,
the $(B,{\vec f})$ and  $(Lk,{\vec r})$ ensembles are equivalent.
This equivalence holds only in the limit of long polymers \cite{ensemble}.   

A twist storing polymer of length $L$ is modelled\cite{bouchiat,nelson} 
as a ribbon $\{{\vec x}(s), {\vec u}(s)\}$
where $s$ , $0\le s\le L$ is an arclength parameter along $\{{\vec 
x}(s)\}$. 
${\vec u}$ represents the   
``ribbon vector'' and is required to satisfy
${\vec u}\cdot{\vec t}=0$
where ${\vec t}=\frac{d}{ds} {\vec x}$.
The ribbon is described as a family of curves
${\vec x}(s)+\epsilon {\vec u}(s),$
where ${\vec x}(s)$ is a curve and
${\vec u}(s)$ represents a slight ($\epsilon$) 
thickening of it along ${\vec u}(s)$ with $\epsilon$ a small parameter.  
Let $\kappa(s) = |\frac{d}{ds} {\vec t}|$ be the curvature 
and $\Omega_3(s)={\vec t}\cdot[{\vec u} \times \frac{d}{ds} {\vec u}],$ 
the
twist of the ribbon. 
In the models we discuss, 
the energy $E({\cal C})$ of a
configuration ${\cal C}$ is given by
\begin{equation}
{E}[{\cal C}]=1/2 \int_0^L ds[A {\kappa^2(s)}+
C {{\Omega}_3^2(s)}],
\label{energy}
\end{equation}
where $A$ is the bending modulus and $C$ the twist modulus.
Other terms can be added as in \cite{nelson} with more
parameters, but these
are not necessary for our purposes. 

In order to compute the partition function $Z({\vec r},Lk)$, at fixed
link $Lk$ and extension ${\vec r}$, we need to specify ${\bf C}$, the 
allowed configurations of the polymer. This is what distinguishes the 
different models we now describe.
In all the models we consider here, the configurations of the 
polymer
are required to obey the following constraints: the polymer extends from  
${{\vec x}(0)}=0$ to ${{\vec x}(L)}={\vec r}$, has fixed tangent vectors 
at the ends, 
${\hat t}(0)={\hat t}(L)={\hat z}$ and 
the ribbon vectors at the ends are given by 
${\hat u}(0)={\hat x}$ and ${\hat u}(L)=R(2\pi Lk){\hat x}$.
$R(2\pi Lk)$ represents a rotation about the $z$ axis through $Lk$ turns. 
In addition, there may be 
further constraints which define the model and such constraints
can alter the {\it topology} of the configuration space and {\it 
qualitatively}
alter the predictions of the theoretical model.
For quantitative agreement with experimental data one needs to take into 
account the geometry and the statistical mechanics of the model. 
 
An important principle to bear in mind is that in Eq. (\ref{part}),
one should only sum over a {\it single} topological class\cite{twist}. 
Thermal agitation
makes the polymer explore different configurations. But
since these agitations only cause continuous changes, the polymer will
remain in a single topological class. We should therefore sum over all
configurations in a {\it single} topological class and not sum over 
distinct
topological classes. Once this principle is understood
and consistently applied, we find that the theoretical picture 
becomes much clearer.

{\bf (a) Worm like chain (WLC):} In this model 
no further constraints are imposed on ${\vec x}(s)$. 
Thus the polymer is allowed to intersect itself. 
While the set of self-intersecting configurations may
be of small measure in the configuration space ${\bf C_a}$, 
such configurations profoundly affect 
the topology of the configuration 
space. For, a WLC polymer 
can release link two units at a time 
by passing through itself\cite{baez,twist,heid}. This ``topological 
untwisting''
results in a collapse of link sectors.  All link values
differing by $2$ 
are in the same topological sector and
one has to sum  $Z({\vec r},Lk)$ over 
$Lk$ classes differing by $2$. This model has been treated in 
(\cite{twist}) and 
leads to a partition function which is periodic in $Lk$ with
period two. 
This {\it periodicity}  is clearly at variance  
with the {\it aperiodicity} seen in the 
experimental curves \cite{strick}, 
where as we mentioned
before, the extension progressively decreases as  
the bead is turned hundreds of times. The discrepancy between  
the WLC model and experimental data 
is clearly due to not taking into account the 
{\it topological} effects of self avoidance. 

However, there is a limit 
(the paraxial worm like chain (PWLC))
of the WLC model which yields an {\it aperiodic}
free energy profile. In the high tension regime, (large force or 
$|{\vec r}|$ comparable to the contour length $L$), the polymer is
essentially straight between its ends and the tangent vector ${\hat t}$
only makes small deviations from the 
direction ${\hat z}$ of the applied force,
which we call the north pole (of the sphere of tangent directions). 
In this regime, one can 
do perturbation theory and 
approximate
the sphere of directions by the tangent plane at the north pole.
This defines the

{\bf (b) Paraxial worm like chain (PWLC):} the tangent vector ${\hat t} = 
\frac{d 
{\vec 
x}}{ds}$ must be near ${\hat z}$, the direction of the applied force 
{\it i.e.} ${\hat t} \sim {\hat z}$.

The PWLC model has been treated in \cite{nelson} and a simple analytic
form for the writhe distribution is given in \cite{writhe}.
In this model, the polymer cannot release link by passing through itself
since the high force prevents the molecule from looping back on itself to 
release link. However, this model only works in the limit of large
forces (theoretically, infinite) and does not address the low force
regime, which is experimentally accessible.
In order to prevent the polymer from releasing link at low forces
by passing through itself, one would like to impose a
condition preventing the polymer from intersecting itself.
This defines the 

{\bf (c) Self-Avoiding Worm Like Chain (SAWLC):}
the configurations must be self avoiding:
\begin{equation} 
{\vec x}(s) \neq {\vec x}(s') ,
\label{sawlc}
\end{equation}
for 
$s \neq s'.$
However, even this condition (\ref{sawlc}) is not sufficient to prevent
topological untwisting. 
The experiments reported in \cite{strick} study the elastic properties of 
{\it linear} DNA molecules, not circular ones.
So we need to address the issue of modelling open ribbons.  
Open ribbons again need careful handling
because of {\it topological
untwisting}: an open ribbon can untwist itself
by two turns
even in model (c)
by going around its end and thus
can change its link $Lk$ by 2.
Consequently, a configuration
with a link $Lk$ is in the same topological class as a configuration
with link $Lk+2$ and 
we once again get a collapse of $Lk$ sectors leading to a periodic
partition function\cite{twist}:
$$Z({\vec r}, Lk+2) = Z({\vec r}, Lk).$$
In real experiments, there {\it is}
an obstruction
to releasing two units of
link by going around the ends: the size of the magnetic bead is
large enough to prevent topological untwisting of the
polymer over the duration of the experiment.
To model this obstruction to topological untwisting
we close the open ribbon with a reference ribbon that goes through
the bead, makes a large fixed circuit and returns to meet the 
polymer at the glass slide.
We require that the reference ribbon and the real ribbon
together form a 
non-self-intersecting
closed ribbon. (Note that we can allow for a
fractional link by permitting a discontinuity in
the ribbon vector ${\hat u}$ at $s=L$:
${\hat u}(L-\epsilon)= R{\hat u}(L+\epsilon)$ where $R$
represents a rotation about ${\hat t}$ and ${\epsilon}$ is 
a small parameter.)
The condition of self avoidance
keeps the real ribbon from going
around its ends and effectively constrains the applied link. It is
understood, that when we sum over polymer
configurations, we only sum over configurations in which the
real polymer is changed.
This is the ``active part'' of the polymer, as opposed to the
``passive'' reference ribbon which is {\it not} summed over.
The self avoiding worm like chain is defined by only
allowing configurations which do not intersect themselves
or the reference ribbon. This configuration space is
called ${\bf C_c}.$

As we mentioned before, this model is not tractable analytically.
Ideally we would like to work with ${\bf C_c}$ and sum
over (the active parts of) all non self-intersecting closed 
ribbons extending from 
${\vec 0}$
to ${\vec r}$ (and returning to ${\vec 0}$ along a reference curve)
with fixed $Lk$. However, the constraint of  
self-avoidance
is far from easy to handle! The constraint is non-local in the 
arc length parameter $s$ and 
is hard to implement analytically\cite{kamien}. 
This has led 
Bouchiat and Mezard \cite{bouchiat} to propose a new model which they
call the Worm like rod chain (WLRC). 

{\bf (d) Worm Like Rod Chain (WLRC) :} is defined by imposing the
additional constraint on the WLC configuration space ${\bf C_a}$ that the
tangent vector ${\hat t}$ must nowhere point south ${\hat t} \neq {-\hat
z}$. Recall that the force is applied in the $z$ direction, which we call
north.
This is also the direction in which the tangent vectors at the end 
are held fixed.   
Unlike the self avoidance constraint, the ``south avoidance''
constraint is local in the arc length parameter $s$. As a result, the
model is amenable to semi-analytic treatment. The authors of
\cite{bouchiat} enforce the constraint by using a repulsive potential at
the south pole of the sphere of directions. The width of this potential is
${\epsilon}$, which serves as a cutoff. This model gives an aperiodic free
energy profile and a very good fit to the experimental data after the
cutoff is suitably adjusted. While the WLRC model does agree with the
data, its theoretical significance is not clear to many workers in the
field (see the recent exchanges in PRL\cite{maggs,mezreply} over the
significance of the cutoff and the local writhe formula).  The purpose of
this paper is to explain the nature of the relationship between the WLC,
the WLRC and the PWLC models, explore different regimes of the elasticity
of twist storing polymers and note that one can continuously interpolate
between them in experimental situations.
To summarize, the self avoiding Worm Like Chain (SAWLC (c)), the paraxial 
worm like chain
(PWLC (b)) and the Worm Like Rod Chain (WLRC (d))
emerge by placing additional constraints on
the Worm Like Chain
WLC model (a)\cite{twist}. 
One important qualitative feature that 
distinguishes WLC from the other models is the 
fact that 
WLC predicts a {\it periodic free energy} whereas the other models 
predict {\it aperiodic free energies}. 
In the WLC model the system is delocalized over 
$Lk$ sectors. This can be visualized in terms of a 
periodic multiple well
potential where the thermal energy is high enough for a particle 
to visit all wells. 
From this point of view SAWLC, WLRC and PWLC are topologically and 
qualitatively similar and markedly different from the WLC model.

The topological differences between these various models
can also be expressed in terms of 
the variable $B$ conjugate to the applied
link $Lk$ variable. 
This conjugate variable $B$, 
can be interpreted as a magnetic field\cite{bouchiat,writhe,twist}. 
In the a) WLC model $B$ has the 
interpretation of the field of a magnetic monopole located at the 
center of the sphere of tangent directions to the polymer\cite{twist}
and the magnetic flux is quantized and this leads to a periodicity
of the free energy.
In the b) PWLC model, the excursions of 
the polymer are confined to the tangent plane at the north pole,
and the magnetic field $B$ is on a plane and therefore not quantised. As a 
result the free energy is not periodic in the link  
$Lk$. 
\vbox{
\epsfxsize=6.0cm
\epsfysize=6.0cm
\epsffile{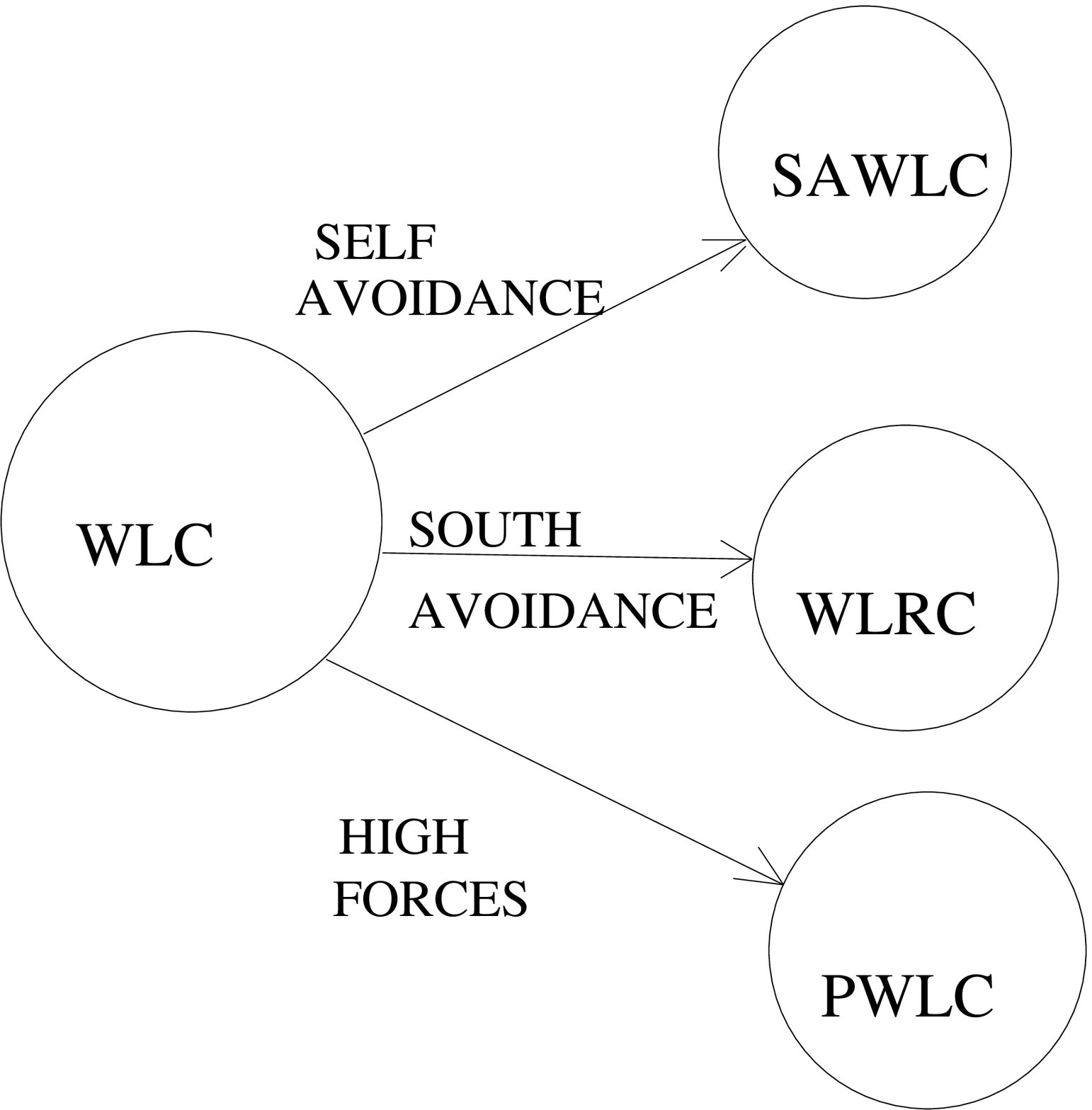}
\begin{figure}
\caption{
The free energy of the WLC model has a periodicity (due to a 
collapse of $Lk$ sectors) which can be removed by imposing
self-avoidance or south-avoidance or large stretch forces. 
All these are different ways of separating the $Lk$ sectors and 
preventing the ``collapse of link sectors''.}
\label{Pz}
\end{figure}}

In the WLRC model, because of the constraint of south avoidance,
$B$ is defined on a sphere punctured at the south pole.
Since the topology of the punctured sphere is identical 
(by stereographic projection) to that of the plane, 
there is no quantization condition just as in the PWLC model
and again, this leads to an aperiodic free energy profile (see Fig.1).

\section{Wreathe and Writhe}
In fact the right model to describe a real polymer under 
a torsional constraint would be c) the self-avoiding WLC (SAWLC)
model which puts a constraint on the polymer passing through itself
or the reference ribbon.
However, as discussed earlier, such a model is hard to solve because
of the non-locality of the constraint. We show below 
that the WLRC model captures some of the essential topological 
features of the SAWLC model and thus enables us to deal with a simpler
and solvable model where the constraint reduces to a {\it local} one 
instead of a 
{\it non-local} one. 
Below, we show that i) the WLRC model captures
the right qualitative behaviour of the Partition function. The WLRC
partition function is {\it aperiodic} in the applied link unlike
the WLC partition function \cite{twist} which is {\it periodic}.
ii) Quantitatively, the WLRC partition function is a better
approximation to the SAWLC partition function than the PWLC partition
function and works even for low forces. 

We begin with the qualitative features, which are of a topological nature.
The central quantity of interest [Eq.{\ref{part}}] is the partition
function $Z(Lk)$ which is the sum over ribbon configurations with fixed
link ($Lk$).  C\u{a}lug\u{a}reanu's theorem\cite{cal} tells us that $Lk 
= Tw +
Wr, $ where the twist $Tw=\int{{\Omega}_3(s) ds}$ is the integral of the
local twist $\Omega_3(s)$ along the curve. Writhe ($Wr$) is a quantity
that only depends on the curve $\{{\vec x}(s)\}$ and not on the ribbon
vector $\{{\vec u}(s)\}$\cite{kamien}. The problem thus neatly splits into
two parts.  The link distribution is the convolution of the twist
distribution and the writhe distribution \cite{bouchiat,writhe}. The twist
distribution is a quadratic path integral and is easily evaluated.  The
problem that remains is to compute the writhe distribution. Thus the
problem reduces from one defined on the space of ribbon 
configurations (${\bf C}$) to
one defined on the space of ribbon backbones (${\bf {\tilde C}}$) or the 
space of curves.  We
can formally set the twist elastic constant to infinity ($C = \infty$,
the molecule is 
impossible to twist) so that $Lk = Wr$ and then the $Lk$ distribution is 
identical to the $Wr$
distribution.

The writhe is a non-local quantity defined only on closed
non-self intersecting
curves:
Let $s$ range over the entire length $L_0$ of the closed ribbon
(real ribbon $+$ reference ribbon) and
let us consider ${\vec x}(s)$ to be a periodic function
of $s$ with period $L_0$.
Let ${\vec
R}(s,\sigma)={\vec x}(s+\sigma)-{\vec x}(s)$. ${\vec R}(s,\sigma)$
is non-vanishing for $\sigma\neq 0,L_0$ and the
unit vector ${\hat R}(s,\sigma)$ is well-defined. It is easily checked
that as $\sigma \rightarrow \{0,L_0\},$
${\hat R} \rightarrow\{{\hat t},{-\hat
t}\}$
respectively. The C\u{a}lug\u{a}reanu-White 
writhe is given by\cite{heid,dennis,footcal}

 \begin{eqnarray}
{\cal W}_{CW}= \frac{1} {4\pi}\oint_{0}^{L_{0}}ds\int_{0+}^{L_{0}-}d\sigma
[\frac{d{\hat{R} (s,\sigma)}}{ds}
\times \frac{d{\hat{R}(s,\sigma)}}{d\sigma}]\cdot{\hat{R}}
 \label{writhe}
 \end{eqnarray}

The non-locality of ${\cal W}_{CW}$ makes it difficult to handle
analytically.
However, the key point to note is that {\it variations} in ${\cal W}_{CW}$
are {\it local} \cite{fuller}.
Let ${\tilde {\cal C}}(\tau)$ be a family of non-self intersecting closed
curves with
writhe ${\cal W}_{CW}(\tau)$.
Taking the ${\tau}$ derivative of Eq. (\ref{writhe}) we find that the
resulting terms can be rearranged to give\cite{ald}

\begin{eqnarray}
\frac{d{\cal W}_{CW}}{d\tau} = \frac{1} {2\pi}\oint_{0}^{L_{0}}
ds [\frac{d{\hat{t} }}{ds}
\times \frac{ d{\hat{t}}}{d{\tau}}]\cdot{\hat{t}} 
\label{dwrithe}
\end{eqnarray}
 which clearly has the interpretation of the rate at which
${\hat t}(s, \tau)$ sweeps out a solid angle in the
space of directions. Note that Eq.(\ref{dwrithe})
is a {\it single} integral \cite{ald} and therefore
a local, additive  quantity. 
Changes of writhe are local integrals and get contributions only
from the active parts of the curve. 
Eq. \ref{dwrithe} can be rewritten as:
\begin{eqnarray}
\frac{d{\cal W}_{CW}}{d\tau} = \frac{1} {2\pi}
\frac{d{\Omega}({\tau})}{d\tau}
\label{dtwrithe}
\end{eqnarray}
where $\Omega$ is the solid angle
enclosed by the oriented curve \{${\hat t}(s)|0\leq s \leq L_0$\}
on the unit sphere of tangent directions
as $s$ goes from
$0$ to $L_0$ \cite{kamien,twist}.
Note that $\Omega$ is
only defined modulo $4\pi$:
for a solid angle $\Omega$ to the left of the oriented curve ${\hat t}(s)$
is equivalent to a solid angle $(4\pi-\Omega)$ to the right.
$d\Omega/d\tau$ is however well defined and local.
Integrating Eq. (\ref{dtwrithe}) we arrive at\cite{fuller}:
\begin{eqnarray}
{\cal W}_{CW}({\tau}) = \frac{1} {2\pi}
{\Omega} -1 +2n
\label{twrithe}
\end{eqnarray}
where $n$ is an arbitrary integer. 

There {\it is} a quantity one can
construct from the writhe which is well defined on {\it all} curves
(not just simple ones)
$$w({\tilde {\cal C}})=exp[i\pi{\cal W}_{CW}({\tilde {\cal C}})]
= -exp[{i\Omega}/2] $$
is a complex number of modulus
unity which we call the
{\it wreathe}. From the geometric phase point of view
(see the analogy developed in \cite{twist}) 
the wreathe is a very natural object to consider: it 
is simply the geometric phase of a spin half system in a 
cyclically varying magnetic
field.
When a curve is passed through itself, ${\cal W}_{CW}$ jumps by $2$, but
the wreathe is unchanged. We can therefore smoothly extend
$w$ to all closed curves {\it including non-simple curves}.
The wreathe can be used to define a local quantity,
the ``Fuller writhe'' \cite{fuller} for curves which
are nowhere south pointing. 

Let us define the ``wreathe angular velocity'' 
${\cal A}_{\tau} = 
-iw^{-1}{\frac{dw}{d\tau}}=\frac{1}{2}d\Omega/d\tau$
as a ``vector potential'' on the space of curves. It is easily seen 
that ${\cal A}_\tau$ is curl free. However, 
${\cal A}_{\tau}$ is not a gradient in the space of {\it all} curves:
there exists no function $W$ defined on {\it all} curves such that
${\cal A}_{\tau}$ is given by $\frac{dW}{d\tau}$. This follows because 
there exist closed circuits ${\tilde {\cal C}}(\tau)$ in the space of 
curves for 
which the integral $\oint d\tau{\cal A}_\tau$ 
is non zero.
Such circuits enclose a nonzero
topological flux and are link changing closed circuits (LCCCs).
They describe the process of topological untwisting.

Let us choose a fiducial curve ${\tilde {\cal C}_*}$, 
which goes from ${\vec 0}$ to ${\vec r}$ in a straight line, whose tangent
vector is identically north pointing. 
Observe that all south avoiding curves are deformable to the fiducial curve
${\tilde {\cal C}_*}$ via south avoiding curves. One simply deforms the 
tangent vector ${\hat t}(s)$
along the unique shorter geodesic connecting ${\hat t}(s)$
to the north pole. We now define the Fuller writhe as
$${\cal W}_F=1/\pi \int_{\tilde {\cal C}_*}^{\tilde {\cal C}} d\tau {\cal A}_\tau-1$$
Writing the unit tangent vector as ${\hat t}=(\sin{\theta}\cos{\phi},
\sin{\theta}\sin{\phi},\cos{\theta})$,
$\int_0^{1}{d{\tau} \frac{d\Omega}{d\tau}}$ can be written as
$\int{ds\frac{d{\phi}}{ds}(1-\cos{\theta})}$ for all curves for which the
tangent vector never points towards 
the south pole of the sphere of tangent directions. We
can therefore write 
a ``local writhe'' on such curves which we call ``south avoiding curves'':
\begin{eqnarray}
{\cal W}_{F} = {\frac{1}{2\pi}}\int{ ds {(1- \cos{\theta})
{\frac{d{\phi}}{ds}}}} -1.
\label{sangle}
\end{eqnarray}
While (\ref{sangle}) is expressed in local co-ordinates on the sphere,
it has a clear geometric meaning:
$2\pi(1+{\cal W}_F)$ is equal to the solid angle swept out by the unique shorter
geodesic connecting the tangent 
vector ${\hat t}$ to the north pole\cite{panch,js}.
This definition is explicitly {\it not} rotationally 
invariant, since it uses a fixed fiducial 
curve ${\tilde {\cal C}_*}$ and singles out a preferred direction.  

This definition of ``Fuller writhe'' is motivated by a theorem of 
Fuller\cite{fuller}, which gives the CW writhe of a curve
which is deformable to the fiducial curve by a family of curves which
is self avoiding {\it and} south avoiding. For such curves, the
Fuller writhe agrees with the CW writhe. However,
in general, the ``Fuller writhe'' is not
equal to the CW writhe, which is what appears in the
C\u{a}lug\u{a}reanu-White formula.
In fact, the definition of Fuller writhe extends
easily to all curves which are south avoiding,
since Eq. (\ref{sangle}) does not require any more conditions.

We remarked earlier that ${\cal A}_\tau$ is not a gradient on the
space of all curves.
Only in certain restricted classes of curves can it be expressed
as a gradient. For example, ${\cal A}$ is a gradient in the space of self 
avoiding
curves  
${\cal A}_{\tau} =\pi 
\frac{d{\cal W}_{CW}}{d\tau}$. It is also a gradient in the 
space of south avoiding curves
${\cal A}_{\tau} = \pi
\frac{d{\cal W}_{F}}{d\tau}$\cite{footprint}. 

To summarise, ${\cal W}_{CW}$ is defined on all self-avoiding curves,
${\cal W}_F$ on
all south avoiding curves. On curves which are 
deformable to the fiducial curve ${\tilde {\cal C}_*}$ through
south {\it and} self avoiding curves the two notions agree \cite{fuller}
(${\cal W}_{CW}={\cal W}_F$). 
When a curve passes through itself,
${\cal W}_{CW}$ jumps by two and when the tangent vector to a curve swings
through the south pole, ${\cal W}_F$ jumps by two units. The writhe is a real
number which has {\it both}
geometric and topological information. The topological part is the
integer part of $\frac{Wr}{2}$ and the geometric part is the
fractional part of $\frac{Wr}{2}$. The geometric part is completely
captured by wreathe but the topology is lost since wreathe
is insensitive to changes in writhe by $2$ units.
From the definitions it is clear that on curves where these quantities
are well defined,
\begin{equation}
-i{\frac{{w}^{-1}}{\pi}}{\frac{dw}{d\tau}}= 
\frac{d{\cal W}_{CW}}{d\tau}=
\frac
{d{\cal W}_{F}}{d\tau} 
\label{change}
\end{equation}
so changes in writhe are the same whether measured by ${\cal W}_{CW},{\cal W}_F$
or $w$. Note that $w$ is well-defined on {\it all} curves.

Let us now apply these general topological ideas to understand the 
configuration space of the polymer ribbon. As stated earlier, we need to
sum over a single topological sector of the configuration space. For a
ribbon, this means summing over a single knot class $K$ of the ribbon
backbone ${\tilde {\cal C}}$ and a single link class $Lk$ of the ribbon. 
So, in fact the 
true partition function $Z_K({\vec r},Lk)$ would depend not only
on the extension and the link, but also the knot class of the ribbon 
backbone. Needless to say, this is a hopelessly intractable problem.

The constraint of self avoidance is a non-local constraint in $s$
and very hard to handle analytically. Even in the pure 
bend model \cite{marko,bend} 
one cannot handle this constraint and so 
one just gives up self avoidance 
and sums over all 
configurations. This is of course an approximation, but it seems to
work well. Instead 
of evaluating (\ref{part}) with ${\tilde {\cal C}}$ equal to simple curves 
in a 
single knot class $K$ (for example, the trivial knot class) :
\begin{equation}
Z_{K} = \sum_{{\cal C} \epsilon K}{e^{-\beta { E}({\cal C})}}
\label{zk}
\end{equation}
we are 
effectively summing over knot classes and computing $Z=\sum_{K}{Z_K}$ 
since we are unable to impose self-avoidance.
The resulting analysis still gives a 
reasonable account of the data\cite{bust,marko}. This suggests that the 
contribution from the non-trivial knot classes may not be significant.
In fact, in the presence of a force $f$, there is an energy 
cost $f L_P$ leading to a suppression $e^{-f L_P}$ of the
probability of forming knots\cite{fenchel}.

In the case of twisting polymers, giving up self-avoidance has a more
serious consequence. Giving up self avoidance leads not only to 
an identification of knot sectors, but also link sectors separated by two 
units. This is due to LCCCs.
These are closed circuits in the space of curves,
which when lifted up to the space of ribbons by continuity
become open and lead to an identification of link sectors.
We show below that LCCCs must pass through both south pointing
and self intersecting curves. 

More precisely, our main result is:
in any closed circuit, the number of signed self crossings is equal to the 
number of signed south crossings.

{\it Proof:} The proof uses the fact that the wreathe is
defined on all curves.
Consider a closed circuit 
${\tilde {\cal C} (\tau)}$ ($0 \leq \tau \leq 1$) in ${\bf {\tilde C}_a}$ 
which 
starts from ${\tilde {\cal C}_*}$ and returns to it: 
 ${\tilde {\cal C}(0)}={\tilde {\cal C} (1)}={\tilde {\cal C}_*}.$
If we now compute the wreathe (which is defined on all curves) we find 
that as $\tau$ varies from $0$ to $1$, $0 \leq \tau \leq 1$, the wreathe
$w(\tau)$ describes a motion on the unit circle and returns to its
starting point. 
The number of times $w(\tau)$ winds 
around the unit circle is given by 
$$k=\frac{-i}{\pi}\int_0^1 d{\tau}w^{-1} dw/d{\tau}
=\frac{1}{\pi}\oint d{\tau} {\cal A}_{\tau} $$
Using Eq. (\ref{change}) we find that 
$k$ measures i) the number of times (counted with sign) the polymer
passes through itself 
(i.e. $k=1/2\int_0^1 d{\tau}\frac{dW_{CW}}{d\tau}$).
ii) the number of times (counted with sign) the 
polymer tangent vector swings through the south pole
(i.e. $k=1/2\int_0^1 d{\tau}\frac{dW_{F}}{d\tau}$).

The above discussion can be ``lifted'' in a sense (made precise in the 
appendix) to the space of ribbons which is the true configuration space. 
Starting with a closed circuit ${\tilde {\cal C} }(\tau)$ and an initial 
ribbon ${\cal C }_{*}$ whose base curve is ${\tilde {\cal C} }_{*}$.
One can lift the circuit ${\tilde {\cal C} }(\tau)$ in the space of 
curves, by continuity to the space of ribbons. However, closed circuits
in the space of curves may be {\it open} in the space of ribbons! This
is the well known anholonomy effect.
Our main result can then be re-expressed on the space of
ribbons:
{\it two ribbons based on the same ribbon backbone ${\tilde {\cal C}_*}$
are homotopic as self avoiding ribbons if and only if they are homotopic
as south avoiding ribbons.}

Proof: The argument above also shows that $2k$ is
equal to the net change in the $Lk$ class of a ribbon if its 
backbone is continuously deformed  along ${\tilde {\cal C}(\tau)}$
(since $Lk=Wr$ in our analysis). 

If we specialise to 
self avoiding curves ${\tilde {\bf C_c}}$ then 
$k = \Delta {\cal W}_{CW}=0$. If we work with south avoiding curves 
${\tilde {\bf C_d}}$, then $k = \Delta {\cal W}_{F}=0$. 
This proves that self and south avoidance present the {\it same
topological obstruction} to release of link $Lk$. 
This provides a formal justification for Bouchiat and 
Mezard's work\cite{bouchiat} where they impose a south avoidance 
constraint in place of a self avoidance constraint. 

$k$ also has the interpretation of the ``quantized magnetic
flux'' passing through the loop \{${\tilde{\cal C}}(\tau); 0\le \tau \le 
1$\}. 
We can write 
$k$ as
\begin{equation}
k=1/(2\pi)\oint_0^1d\tau \oint_0^{L_{0}} ds [\frac{d{\hat{t} }}{ds}
\times \frac{ d{\hat{t}}}{d{\tau}}]\cdot{\hat{t}} 
=\frac{1}{\pi}\oint d{\tau}{\cal A}_{\tau} 
\label{degree}
\end{equation}
which measures the ``topological flux'' passing through ${\tilde {\cal 
C}}$. $k$ is also the degree of the map ${\hat t}(\tau,s)$ from the
torus \{$0\le s\le L_0,0\le \tau \le 1$\} to the sphere
of tangent directions. Clearly if the 
flux is non zero, then the degree is non zero and the tangent vector
must point south somewhere. Removing the south pole \cite{footpun}(as 
Bouchiat and Mezard\cite{bouchiat} do)
forces the degree to be zero and prevents link collapse.

Our arguments above show that the topology of self
avoidance is captured by south avoidance. This is a qualitative
feature related to topology. At this qualitative level, the excised
point on the sphere of tangent directions need not be the south pole
(recall that the force direction is taken to be the north pole), but can
be any direction ${\hat n}$ on the sphere. However, in order to achieve
quantitative success in explaining the experimental data, we show below 
that it is advantageous to choose ${\hat n}$ along the south pole.
Ideally we would like to compute the partition function of the SAWLC model
to compare with experiment. The replacement of SAWLC by WLRC seems to 
give a reasonable account of the data. Why does it work? To see why,
consider the region ${\cal R}$ in the space of curves which can be reached
from the fiducial curve ${\tilde {\cal C}_*}$ without ever self 
intersecting or pointing south. Let us notice the following:
i) The region ${\cal R}$ has a finite measure in the space of curves.
ii) All over ${\cal R}$, ${\cal W}_{CW}={\cal W}_{F}$.
iii) The region ${\cal R}$ dominates the partition function.
More precisely, regions outside ${\cal R}$ are suppressed by a factor
$e^{-f_{eff} L_P}$. Here $f_{eff}=f-B^2/(4L_P)$ is an 
effective force \cite{nelson,writhe} that
takes into account the competition between the stretch and the twist.
Consequently, the WLRC partition function is a good approximation to the
SAWLC partition function over a range of parameter space.

Let us successively consider the regimes of low, intermediate and high
twist. Quantitatively, this is measured by $f_{eff}$.
(i) For $f_{eff}L_{P}$ large, the tangent vector makes small excursions about 
the north pole and $<\theta^2>=1/\sqrt{f_{eff}L_{P}}$ is small. In this
regime, one can approximate the sphere by its tangent plane at the north
pole and as explained earlier, this is the regime of the PWLC model.
(ii) For $<\theta^2>$ of order $1$,($f_{eff} L_P$ of order $1$), this is not
a good approximation, since the geometry of the sphere is not well 
approximated by the planar geometry. However, in this regime, the WLRC 
works well since the suppression factor $e^{-f_{eff} L_P}$ is small. 
Since the WLRC retains the geometry of the sphere (changing only the 
topology, by removing a point), the WLRC is an improvement on the PWLC
model.
(iii) At very low (or negative) $f_{eff}$, in the SAWLC model, the polymer
accommodates writhe by winding around itself. A similar behaviour
occurs in the WLRC model where the 
tangent vector is mostly near the south pole. In both models a cutoff set 
by the thickness of the polymer (see 
below) is needed to regularise the model.

Had we chosen to exclude some other direction
instead of the south pole, the suppression factor would have been 
$e^{f_{eff}L_P\cos{\theta}}$, where $\theta$ is the polar angle of the 
excluded 
point. Clearly, the advantageous choice is $\theta=\pi$ {\it i.e.}, the 
south pole of the sphere. 
The choice ${\theta = \pi}$ leads to an axially symmetric quantum 
mechanical problem, for which the Hamiltonian can be solved. 
This is exactly what Bouchiat and Mezard choose
in their WLRC model.

{\bf Need for a cutoff:}
In summing over all south avoiding paths one finds that the 
problem of computing Eq. (\ref{part}) is ill defined. 
Recall that ${\tilde {\bf S_d}}$, the set of south pointing paths 
have been removed from ${\tilde {\bf C_a}}$, the WLC configuration space
to get the WLRC configuration space 
${\tilde {\bf C_d}}= {\tilde {\bf C_a}}-{\tilde {\bf S_d}}.$
The problem is that near ${\tilde {\bf S_d}}$ there are paths
(points in ${\tilde {\bf C_d}}$) with arbitrarily large writhe and 
vanishingly small energy. This problem is due to paths which 
wind around the south pole in tiny circles accumulating 
large writhe at zero energy cost \cite{comtet}. 

This problem does not occur at large forces. At large forces there is a
large energy cost preventing the tangent vector from visiting the south
pole. Indeed, in the paraxial limit there is no pathology
\cite{maggs,nelson,writhe}. As one lowers the force, the energy cost for
accommodating writhe tends to zero near ${\tilde {\bf S_d}}$ and these
dominate the sum in Eq. (\ref{part}).  In order to get sensible results
from the WLRC model one has to impose a cutoff: the tangent vector is
excluded from a small circle $ \theta= \theta_c = \pi - {\epsilon} $
around the south pole.  In more detail, we see that for paths winding
around the south pole the energy per unit length $E/L \sim {\epsilon}^2$
while the writhe per unit length $Wr/L \sim 2 -{\epsilon}^2$. This implies
that in the limit ${\epsilon} \rightarrow 0$ any amount of writhe can be
accommodated at zero energy cost.  Maggs and Rossetto\cite{maggs} claim
that this pathology of the WLRC model is an artefact of the {\it local}
Fuller writhe formulation. However, we notice that the {\it same}
pathology afflicts the SAWLC model (which involves a {\it non local}
notion of writhe, the CW writhe) as well. This can be seen as follows.  
Consider a polymer twined on a cylindrical rod \cite{footcww} like a
garden climber twines around a pole. 
As explained in Fuller\cite{fuller},
for a plectoneme of pitch angle $\alpha$,
(in the limit that $\alpha$ goes to $\pi/2$
the helix is almost
a straight line)
the curvature goes as $\kappa=(d/2)^{-1} cos^2 \alpha$, 
while the writhe per 
unit length ${\cal W}_{CW}/l$ goes as $sin\alpha$, where $d$ is the
diameter of the rod.
(See Fuller, figure 1 and its caption in the 1971 paper).
From the geometry $cos \alpha =\pi d/(l)$, where $l$ is the length
of the polymer. 
When $\alpha$ goes to $\pi/2$, the 
writhe (per unit length) goes to a
constant and the energy per unit length goes as $d^2$. Note that we
are holding the length of the polymer constant as we take the
limit $\alpha$ going
to $\pi/2$.

We thus come to the conclusion that {\it both} SAWLC (involving 
the {\it non local} CW writhe) and WLRC (involving the {\it local} Fuller
writhe) suffer from the {\it same} pathology which can be cured 
by introducing a cutoff parameter which has the {\it same} origin
( the finite thickness of the polymer) in both formulations. 
In fact Bouchiat and Mezard's WLRC model is
a simple analytically tractable model which captures the essential 
qualitative physics of 
the SAWLC model. 
 
\section{Tunnelling}

So far we have used topological ideas to split up the configuration space
into disconnected sectors. 
However, the difference between topology
and energetics can be blurred by activated processes\cite{maggs}. To 
appreciate this
point, consider a Brownian particle 
in a double well potential. If the 
barrier height is not too large, the particle will randomly visit both
wells during the observation time. If one increases the 
barrier height, the 
activated process is exponentially suppressed. In the limit that the
barrier height goes to infinity, we may say that the two wells belong
to distinct topological sectors and sum over only one of them. The
transition from one regime to the other is measured by the Kramers' 
time $\tau_{\rm Kramers} \sim \exp(V/kT)$, where $V$ is the height
of the barrier and $kT$ the temperature. If $\tau_{\rm Kramers}>>\tau_{\rm 
Observation}$, then we would sum over a single well, but if the reverse is
true, then we would sum over both wells. By tuning $V$, the barrier 
between 
the wells, we can continuously interpolate 
between the single well and the double well descriptions. 
The picture can be extended to include multiple wells,
for example, a potential profile $V(x)= -V_0 Cos x$ which has 
multiple minima at $x= 2 n \pi$.
In the polymer context, the different $Lk$ sectors correspond
to different minima of this potential.
Tunnelling between different minima is an activated process
and the classical solution that dominates this process is an
``instanton'' \cite{rajaraman}. The physics involved here is
very similar to the tunnelling between different topological
sectors of vacuum classical configurations in QCD. The ``$\theta$
vacuaa'' that emerge in QCD are delocalised over all the topological
sectors, just as the WLC polymer is delocalised over link sectors. 
Tunnelling between different minima is ruled out in the 
models (b), (c) and (d) but permitted in the model (a). 
In order to interpolate between these models theoretically we can 
soften the cutoff in the WLRC model by putting a potential of 
height $V_0$ and width $\epsilon$. As $V_0 \rightarrow \infty$ we recover
the WLRC model and as $V_0 \rightarrow 0$ we recover the WLC model. 
$V_0$ serves as a natural parameter in interpolating 
continuously between these models. 
In the next section we show how one can experimentally explore the 
continuous transition between these models
using the catalytic effect of enzymes.  
We will see how addition of Topoisomerase II enzymes effectively 
lowers the energy barrier $V_0$ permitting the molecule to 
pass through itself and thus enabling the polymer to be 
delocalised over distinct link sectors.  

\section{Proposed Experiments}
We now explore the experimental realization of the WLC model (model (a)). 
The process of ``topological untwisting'' can be
made possible by using the enzyme Topoisomerase II to permit the 
DNA molecule to pass through itself\cite{brown,goldstein}. 
(Similar experiments are referred to in \cite{bauer}.)
As the concentration of enzyme $c$ is increased we would expect a 
transition from aperiodic behavior to periodic WLC behavior.
A low concentration would correspond to a large energy barrier to passing 
through itself and a high concentration lowers the energy barrier. We 
may expect the energy barrier $V_0$ to go as $c^{-1}$. 
A similar effect is also expected to happen 
as a function of force $f$. At 
large forces (or extensions), the energy barrier to looping back and
passing through itself is prohibitive. As one lowers the force, one
expects again a transition from the aperiodic behavior noticed in 
a paraxial worm like chain (PWLC) model Ref.[\cite{nelson}, \cite{writhe}) 
to the periodic WLC behavior \cite{twist}. 
The height of the barrier can be worked out in the 
WLRC model ( model (d)).  It is given by $V_0+2f$ 
where $V_0$ is 
the 
repulsive potential at the south pole of the sphere of tangent 
directions and $f$ is the applied stretching force. 
\vbox{
\vspace{1.5cm}
\epsfxsize=6.0cm
\epsfysize=6.0cm
\epsffile{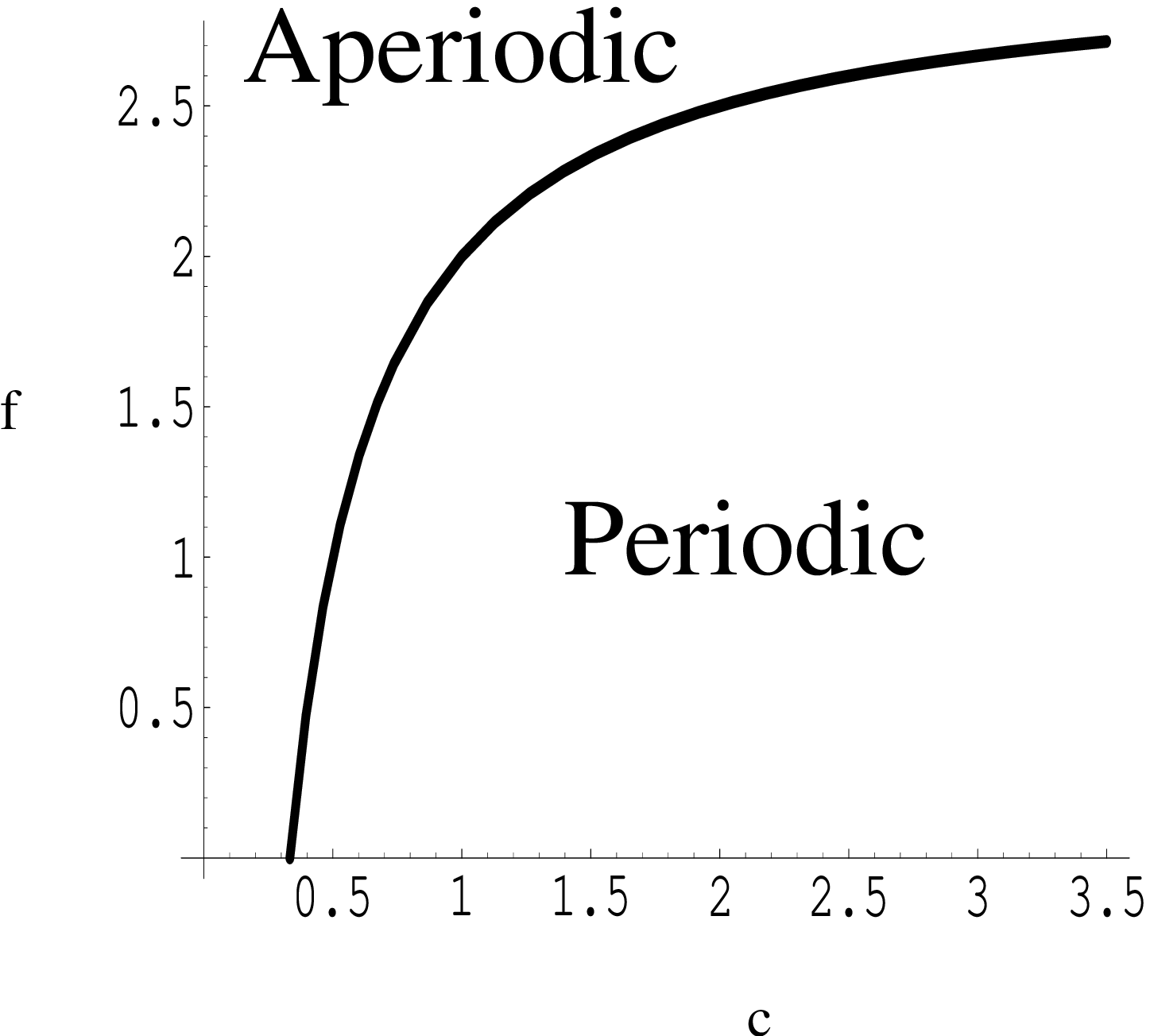}
\begin{figure}
\caption{
The theoretically expected ``phase diagram''
in the force-concentration ($f-c$) plane. $c$ is the concentration 
of the enzyme Topoisomerase II, which permits self crossing.}
\label{Phase}
\end{figure}}
Plotting the lines of constant Kramer's time gives a qualitative 
phase diagram in the force-concentration ($f-c$) plane (see Fig.2). 
At high forces and low concentrations the free energy is aperiodic. 
At low forces and high concentrations the free energy is periodic in 
the link.

During the process of replication the DNA molecule undergoes supercoiling
and it needs to unwind to release its stress. The
viscous cell environment offers resistance to this process. 
The unwinding of these supercoiled structures takes place via 
Topoisomerase enzymes which cut the polymer and helps it to  
release its stress. This real 
biological context is where
the WLC periodic free energy finds a natural application. These
effects can be studied under controlled circumstances in 
single molecule experiments.

\section{Conclusion}

In this paper we have provided a formal justification for the WLRC model
in terms of the topological effects of self-avoidance. 
Our main result is that:
In any closed circuit of curves, the number of signed self-crossings
is equal to the number of signed south-crossings.
Or equivalently, in the language of ribbons,
{\it two ribbons based on the same ribbon backbone ${\tilde {\cal C}_*}$
are homotopic as self avoiding ribbons if and only if they are homotopic
as south avoiding ribbons.}

Replacing the non-local 
notion of self-avoidance by the local notion of south
avoidance results in the analytically tractable WLRC model.
The C\u{a}lug\u{a}reanu-White formula for the writhe 
is explicitly nonlocal (the formula
is expressed as a double integral) and no local formula valid on all 
simple curves exists. In Fuller's treatment\cite{fuller} a theorem
is proved that the writhe difference for two closed non self-intersecting 
space curves is given by a local formula 
under certain conditions \cite{fuller,fain}.
Taking the special case that
the reference curve has constant tangent vector ${\hat z},$ we arrive
at the formula [Eq. (\ref{sangle})] for non self-intersecting
(i.e simple) curves which are nowhere south
pointing. Note, however, that the restriction of non self-intersection can be 
relaxed since it nowhere appears in the Fuller writhe formula. Our 
treatment which 
uses the wreathe as the starting point has a larger domain of validity 
since wreathe is well defined on {\it all} curves.
Our main result provides the theoretical framework and justification
for the calculational scheme employed by \cite{bouchiat}
in the WLRC model.

We have focussed on long polymers to keep this discussion simple and
because the experimental situation permits it. Needless to say,
the general topological discussion applies to short polymers as well. In 
particular consider a very rigid polymer $(\L_p \rightarrow \infty, 
L/L_p \rightarrow 0)$
with the tangent vector clamped at both ends to ${\hat z}$. This system 
can be modelled as a paraxial worm like chain
(even if the applied force is zero) and according to the general 
discussion, will have an aperiodic free energy. In this case, the
energy barrier to looping around is given by the elastic energy
of the stiff polymer and goes as $A/L$. 

We have emphasized the need for introducing a regularizing cutoff in a 
theoretical analysis of this problem.
It has been suggested \cite{maggs} that the divergence in the writhe 
distribution 
seen in Ref. \cite{bouchiat} is an artefact of the Fuller formulation of 
the writhe
and will not be present in  a CW formulation. We have shown that the
CW writhe also suffers from the same pathology. 
One sees \cite{fuller} that it costs (almost) no energy to accommodate
any amount of writhe. As a result, large values of writhe are possible. This 
pathology is only cured by considering the physical thickness of DNA
(about $2$ $nm$), which results in an energy cost for writhe and 
suppresses
large writhe fluctuations. Writhe fluctuations can be suppressed either
by increasing the tension on the molecule (PWLC model) or 
by taking into account the physical 
thickness of the molecule (WLRC model).

Ribbons homotopic as self avoiding ribbons are also homotopic
as south avoiding ribbons.
Note that the 
converse is not true: knots cannot be undone 
if one imposes self avoidance. But one can remain 
within the set ${\bf C_d}$ of south avoiding curves and undo any knot. 
Consider 
projecting the knot
onto the $x-y$ plane and by small movements of the tangent vector to 
the polymer undo the knot by passing the ribbon 
through itself since there is no self avoidance constraint. 
When we sum over all south avoiding ribbons we are automatically summing
over all knot classes of the central curve. This is an 
approximation valid only when the contribution of the nontrivial knot
sectors is negligible. 

In future it would be interesting to address the theoretical issues
that stem from considering distinct knot classes. 
The recent class of experiments\cite{quake} probing 
the dynamics of complex knots on single molecules 
of DNA would perhaps enable us to connect the theoretically 
challenging energetic
and topological aspects of knots\cite{fenchel,baez} to experiments.   
One can also experimentally explore the low force stretching 
regime to explore deviation from the pure bend WLC model 
of Ref. \cite{marko} to probe the effect of knotting of a DNA molecule. 

As an offshoot of the observation of the topological distinction between
various models of twisting polymers we are led to a new class of 
experimentally testable predictions. We hope this work will generate
interest amongst experimentalists to test these predictions.

{\it Acknowledgements:}
It is a pleasure to thank D. Bensimon, Y. Rabin, A. C. Maggs
and A. Dhar for discussions. We thank the referee for critical
comments which have improved the paper.
We also thank Harish-Chandra Research 
Institute for the wonderful working atmosphere in which this paper
was written.
\section{Appendix: Ribbons, Curves and Fibres}

To clearly put across the topological point at issue here we
use some mathematical notation.
Let ${\bf C}$ be the configuration space of all closed ribbons
and ${\tilde {\bf C}}$ the space of all curves. There is a natural map
$P: {\bf C} \rightarrow {\tilde {\bf C}}$, which maps each ribbon ${\cal C}$
to its backbone curve ${\tilde {\cal C}}$.
The ribbons based on ${\tilde {\cal C}}$ are classified upto
homotopy by their link class Lk. 
The structure $({\bf C},
{\tilde{\bf C}}, P)$ constitutes a fibre bundle.
Given a closed circuit in ${\tilde {\bf C}}$, which
starts from and returns to ${\tilde {\cal C}}_*$ and a starting ribbon
${\cal C}_*$, one can continuously ``lift'' the circuit
to the space of ribbons.
However, a closed circuit in ${\tilde {\bf C}}$ may lift to an
open circuit in ${\bf C}$.
The initial and final ribbons have the same backbone curve, but
belong to different link  classes.
We refer to such circuits as link changing
closed circuits (LCCCs) in the text.

Let ${\cal C}_*$ be a fiducial ribbon whose backbone
is ${\tilde {\cal C}_*}$. Our sum in
Eq. (\ref{part}) extends only over
the component
connected to ${\cal C}_*$. We call this component $\pi_0^*({\cal {\bf 
C}})$.
In order to perform the sum correctly we need to understand the structure
of $\pi_0({\cal  {\bf C}})$ and $\pi_0^*({\cal {\bf C}})$ correctly in the 
different models.

We start with the model
${\bf (c)}$ the SAWLC model, which is the one of
experimental
interest. $\pi_0({\cal {\bf C}_{\bf c}})$ is labelled by $K$ the knot 
class
of the ribbon backbone $\tilde {\cal C}$ and $Lk$, the link class of the
ribbon based on ${\tilde {\cal C}}$.
(Strictly speaking we should write $[K]$ and $[Lk]$, the
equivalence classes being denoted by the square brackets, but we do not
do so here).
The Link is an integer and
counts the Gauss linking number of the two edges of the closed
ribbon.
Thus $\pi_0({\cal {\bf C}_{\bf c}}) = (K, Lk)$.
$\pi_0^*({\cal {\bf C}_{\bf c}})$, the component connected to ${\cal C}_*$
consists of all ribbons in $(K_*, Lk_*)$, those in the knot class of
${\tilde {\cal C}_*}$ (the unknot class) and which
have the same link as ${\cal C}_*$.

(a) WLC model: If one gives up self avoidance and works with $\bf C_a$,
then the structure of $\pi_0({\cal {\bf C}}_{\bf a})$ is quite
different from $\bf C_c$.
There are no knot classes since any knot can be undone by self crossings.
The even link classes get identified with one another and similarly
the odd link classes. This is due to the presence of
Link Changing Closed Circuits (LCCCs) in $\tilde {\bf C}_a$.
These curves are closed circuits in $\tilde {\bf C}_a$. If one lifts
them up by continuity to $\bf C_a$, the space of ribbons,
we find that the continuous lift of a closed curve in ${\tilde {\bf C_a}}$
may be open in $\bf C_a$. The LCCCs can be recognised by the
fact that they enclose a nonzero topological flux $k=1/\pi \oint d\tau
{\cal A}_\tau$. The final ribbon has a link which differs from the initial
one by $2k$. Thus, the link classes of like parity are identified with each
other and $\pi_0({\bf C_a})$ has just two components. $\pi^*_0({\bf C_a})$
includes all the link sectors which differ from ${\cal C}_*$ by
$2k$. This
is what we refer to as the ``collapse of link sectors''. The configuration
space ${\bf C_a}$ has a vastly different
topological structure from ${\bf C_c}$, the space
of interest. This is where the south avoiding WRLC model ${\bf(d)}$
comes in.  Knots can be undone in ${\bf {\tilde C}_d}$ but (as we see below)
links cannot. So $\pi_0({\bf C_d})=Z$, where $Z$ is the set of integers,
the link $Lk$. The claim that south avoidance captures the topological
effects of self avoidance is based on the following observation. Consider
a LCCC with flux $k$. This provides a continuous deformation in ${\bf 
C_a}$,
between ribbons (based on the same backbone), which differ in $Lk$ by
$2k$. It is shown in the text that these LCCCs pass through self intersecting
curves {\it and} south pointing curves $k$ times
(both counted with sign). Imposing
self avoidance eliminates these LCCS. Alternatively we can
permit self intersections and impose south avoidance. This also has
the same effect of eliminating LCCCs and preventing the collapse
of link sectors, which occurs in ${\bf C_a}$. This proves that
self and south avoidance both present the same topological obstruction
to link release. Mathematically, the link class bundle over ${\tilde {\bf C_a}}$
is nontrivial. 
One can locally trivialise this bundle by omitting
points (in fact sets of measure zero)
from ${\tilde {\bf C_a}}$ . Restricting the bundle to ${\tilde 
{\bf C_d}}$  and
${\tilde {\bf C_c}}$ gives two different trivialisations of the
same bundle.


\begin{references}
\bibitem{bust} C. Bustamante et al,
Current Opinion in Structural Biology {\bf 10}, 279 (2000).
\bibitem{strick}
T. R. Strick et al, {\it Science}
{\bf 271}, 1835 (1996);
D. Bensimon and V. Croquette (private communication)
\bibitem{bouchiat}
C. Bouchiat and M. Mezard, {\it Phys. Rev. Lett.}
{\bf 80}, 1556 (1998).
C. Bouchiat and M. Mezard, Eur. Phys. J. E {\bf 2}, 377 (2000).
\bibitem{maggs}
V. Rossetto and A. C. Maggs, {\it Phys. Rev. Lett.}
{\bf 88}, 089801-1 (2002).
A. C. Maggs and V. Rosetto, {\it Phys. Rev. Lett.} {\bf 87}, 253901 
(2001);
V. Rossetto and A. C. Maggs, {\it J. Chem. Phys.} {\bf 118}, 8864 (2003)
\bibitem{mezreply}
C. Bouchiat and M. Mezard, {\it Phys. Rev. Lett.}
{\bf 88}, 089802-1 (2002). 
\bibitem{cal}
G. C\u{a}lug\u{a}reanu, {\it Czechoslovak Mathematical Journal}
{\bf 11}, 588 (1961); See also
J. H. White {\it Am. J. Math} {\bf 91}, 693 (1969).
\bibitem{fuller}
F.B. Fuller, {\it Proc. Nat. Acad. Sci.} USA {\bf 68}, 815 (1971);
{\it Proc. Nat. Acad. Sci.} USA {\bf 75}, 3557 (1978).
\bibitem{fain}
B. Fain, J. Rudnick and S. Ostlund, Phys. Rev. {\bf E 55}, 7364 (1997).
\bibitem{footwlc}
Note that we use the term WLC model to describe a polymer with bend
{\it as well as twist} degrees of freedom. 
\bibitem{bryant}
Z. Bryant et al., Nature {\bf 424}, 338 (2003).
\bibitem{vino} 
H. B. Gray, W. B. Upholt, J. Vinograd, {\it J. Mol. Biol.} {\bf 62}, 
1 (1971).
\bibitem{nelson}
J. D. Moroz and P. Nelson, {\it Macromolecules} {\bf 31}, 6333 (1998). 
\bibitem{writhe} S. Sinha, 
{\it Phys. Rev. E} {\bf 70 }, 011801 (2004).
\bibitem{kamien} R. D. Kamien, {\it Reviews Of Modern Physics} 
{\bf 74}, 953 (2002); J. D. Moroz and R. D. Kamien, 
{\it Nucl. Phys. B} {\bf 506}, 695 (1997);
W. Kung and R. D. Kamien, 
{\it Europhys. Lett.} {\bf 64}, 323 (2003).
\bibitem{ensemble}
S. Sinha and J. Samuel, {\it Phys. Rev. E} {\bf 71}, 021104 (2005).
\bibitem{twist}
J. Samuel and S. Sinha, 
{\it Phys. Rev. Lett.} {\bf 90 }, 098305 (2003).
\bibitem{baez}
J. Baez and R. Dandoloff, {\it Phys. Lett. A} {\bf 155}, 145 (1991).
\bibitem{heid}
G. H. M. van der Heijden, M. A. Peletier, R. Planqu\'e,
arXiv:math-ph/0310057 (2003).  
\bibitem{dennis}
M. R. Dennis and J. H. Hannay,
arXiv:math-ph/0503012 (2005).
\bibitem{footcal}
Notice that the integrals for the writhe formula are incorrectly given in 
\cite{heid} as two cyclic integrals (i.e. the integration ranges over a 
torus.) The range of integration is in fact a cylinder rather than a 
torus. Our limits are the same as those used in \cite{dennis}.
\bibitem{ald}
J. Aldinger, I. Klapper, M. Tabor, {\it Journal of Knot Theory
and Its Ramifications} {\bf 4}, 343 (1995).
\bibitem{panch}
S. Pancharatnam, Proc. Ind. Acad. Sci. {\bf A44}, 247 (1956).
\bibitem{js}
J. Samuel and R. Bhandari, {\it Phys. Rev. Lett.}
{\bf 60}, 2339 (1988).
\bibitem{footprint}
The space of south avoiding curves is contractible unlike the space of 
self avoiding curves which splits into distinct knot class sectors. 
\bibitem{marko}
J. Marko and E. D. Siggia, {\it Macromolecules} {\bf 28}, 8759 (1995).
\bibitem{bend}
J. Samuel and S. Sinha, 
{\it Phys. Rev. E} {\bf 66 }, 050801(R)(2002).
\bibitem{fenchel}
W. Fenchel, {\it Math. Ann.} {\bf 10} 238 (1929).
\bibitem{footpun}
or for that matter {\it any} point on the sphere. 
\bibitem{comtet} 
A. Comtet, J. Desbois and C. Monthus, {\it J. Stat Phys.} {\bf 73}, 
433 (1993).
\bibitem{footcww} 
In actual terms the polymer wraps around itself to form plectonemic 
structures as in telephone cords. Thus the thickness of the polymer
plays the role of the diameter of the cylinder. 
\bibitem{rajaraman}
R. Rajaraman {\it Solitons and Instantons} (North Holland, Amsterdam 
(1987).
\bibitem{brown}
P.O. Brown and N. R. Cozzarelli, Science {\bf 206}, 1081 (1979);
T. R. Strick et al, Nature (London) {\bf 404}, 901 (2000).
\bibitem{goldstein}
R. E. Goldstein, T. R. Powers and C. H. Wiggins, {\it Phys. Rev. 
Lett.} {\bf 80}, 5232 (1998). The phenomenon of ``Topological Untwisting''
\cite{brown} mentioned here 
is to be contrasted with geometric untwisting \cite{goldstein} 
which involves a continuous release of twist through bending modes 
of a ribbon with a free tangent vector at the end.
\bibitem{bauer} W.R. Bauer, F.H.C. Crick and J.H. White, {\it
Scientific American} {\bf 243},118 (1980).
\bibitem{quake} X. R. Bao, H. J. Lee and S. R. Quake, {\it Phys. Rev. 
Lett.} {\bf 91}, 265506 (2003).
\end{references}
\end{document}